\documentclass{plain}
\usepackage{graphicx}
\usepackage{supertabular,lscape,epsfig}
\usepackage{amssymb}
\usepackage{amsmath}
\usepackage{wasysym}
\SetPages{1}{0}

\SetVol{0}{2023}

\begin{document}

\begin{Titlepage}
\Title{Search for planets in hot Jupiter systems with multi-sector TESS photometry. IV. Null detections in 12 systems}
\Author{G.~~M~a~c~i~e~j~e~w~s~k~i, ~ J.~~S~i~e~r~z~p~u~t~o~w~s~k~a, ~ J.~~G~o~l~o~n~k~a}
{Institute of Astronomy, Faculty of Physics, Astronomy and Informatics,
         Nicolaus Copernicus University in Toru\'n, Grudziadzka 5, 87-100 Toru\'n, Poland,
         e-mail: gmac@umk.pl}

\Received{September 5, 2023\\accepted for publication in Acta Astronomica vol. 73}
\end{Titlepage}

\Abstract{We present the results of our search for nearby planetary companions of transiting hot Jupiters in 12 planetary systems: HAT-P-24, HAT-P-39, HAT-P-42, HAT-P-50, KELT-2, KELT-15, KELT-17, WASP-23, WASP-63, WASP-76, WASP-79, and WASP-161. Our analysis was based on multi-sector time-series photometry from the Transiting Exoplanet Survey Satellite and precise transit timing data sets. We detected no additional transiting planets down to the 2--4 Earth radii regime. For 10 hot Jupiters, no departure from linear transit ephemerides was observed. Whilst we refute long-term variations of the orbital period for WASP-161~b, which were claimed in the literature, we notice a tentative hint of the orbital period shortening for WASP-79~b. In addition, we spot a short-period transit timing variation for KELT-2A~b with the characteristics typical of the so-called exomoon corridor. We conclude, however, that further observations are required to confirm these findings.}{Hot Jupiters -- Stars: individual: HAT-P-24, HAT-P-39, HAT-P-42, HAT-P-50, KELT-2, KELT-15, KELT-17, WASP-23, WASP-63, WASP-76, WASP-79, WASP-161 -- Planets and satellites: individual: HAT-P-24~b, HAT-P-39~b, HAT-P-42~b, HAT-P-50~b, KELT-2~b, KELT-15~b, KELT-17~b, WASP-23~b, WASP-63~b, WASP-76~b, WASP-79~b, WASP-161~b}

%%%%%%%%%%%%%%%%%%%%%%%%%%%%%%%%%%%%%%%%%%%%%%%%%%%%%%%%%%%%%%%%%%%%%%

\section{Introduction}

Our better understanding of planetary configurations with massive planets on tight orbits (hot Jupiters) lends insight into many aspects of the formation and evolution of planetary systems. Statistical studies show that hot Jupiters are relatively rare (e.g., Fressin \etal 2013, Bryant \etal 2023), and in overwhelming cases, they are devoid of nearby planetary companions (e.g., Hord \etal 2021). However, the sample of hot Jupiters in compact planetary systems is constantly growing (see Sha \etal 2023 for a recent review), posing a challenge for theoretical scenarios of the origin of hot Jupiters. 

This paper continues our attempts to detect additional planets in a blind sample of systems with hot Jupiters (Maciejewski 2020, 2022, Maciejewski \etal 2023). Our objectives are achieved by searching for transit signals in photometric time series from the Transiting Exoplanet Survey Satellite (TESS, Ricker \etal 2015) and by analysing transit timing data, in which effects of gravitational perturbations acting from unseen companions could be hidden. 

Here, we report on the results obtained for 12 systems: HAT-P-24, HAT-P-39, HAT-P-42, HAT-P-50, KELT-2, KELT-15, KELT-17, WASP-23, WASP-63, WASP-76, WASP-79, and WASP-161. Although our analysis did not detect additional planets, we revised these systems' parameters and found some hints for transit timing variations for the planets KELT-2A~b and WASP-79~b.

%%%%%%%%%%%%%%%%%%%%%%%%%%%%%%%%%%%%%%%%%%%%%%%%%%%%%%%%%%%%%%%%%%%%%%

\section{Systems of the sample}

Tables 1 and 2 contain the investigated systems' observational, stellar, and planetary properties. Below, we provide a comprehensive summary of literature records found for each system.

%%%%%% TABLE 1

\MakeTable{lcccccc}{12.5cm}{Observational properties of the systems of the sample} 
{\hline
System  & RA (J2000)  & Dec (J2000) & $m_{\rm G}$  & Distance & $d_{\rm tr}$      & $\delta_{\rm tr}$ \\
        & hh:mm:ss.s  & $\pm$dd:mm:ss.s   & [mag]    & [pc]     & [min]             & [ppth]            \\
\hline 
HAT-P-24 & 07:15:18.0 & +14:15:45.4  & 11.64 & $413.2\pm2.5$ & $302^{+11}_{-10}$    & $10.13^{+0.18}_{-0.21}$ \\
HAT-P-39 & 07:35:02.0 & +17:49:48.2  & 12.22 & $706.6\pm7.0$ & $252^{+20}_{-16}$      & $10.52^{+0.29}_{-0.33}$ \\
HAT-P-42 & 09:01:22.6 & +06:05:50.0  & 11.99 & $413.9\pm4.3$ & $231.0^{+8.8}_{-12.8}$ & $ 6.64^{+0.18}_{-0.15}$ \\
HAT-P-50 & 07:52:15.2 & +12:08:21.8  & 11.68 & $491\pm11$    & $222^{+16}_{-14}$      & $ 6.22^{+0.18}_{-0.20}$ \\
KELT-2   & 06:10:39.3 & +30:57:25.7  &  8.59 & $134.5\pm0.4$ & $302^{+11}_{-10}$      & $ 4.48 \pm 0.05$ \\
KELT-15  & 07:49:39.6 & --52:07:13.6 & 11.07 & $316.6\pm1.0$ & $240.2^{+6.0}_{-4.9}$  & $ 9.41^{+0.09}_{-0.10}$ \\
KELT-17  & 08:22:28.2 & +13:44:07.1  &  9.22 & $227.8\pm1.1$ & $206.6^{+4.0}_{-3.5}$  & $ 8.53^{+0.09}_{-0.08}$ \\
WASP-23  & 06:44:30.6 & --42:45:42.7 & 12.31 & $205.8\pm0.4$ & $145.7^{+5.7}_{-4.9}$  & $17.79^{+0.50}_{-0.54}$ \\
WASP-63  & 06:17:20.7 & --38:19:23.8 & 10.92 & $288.9\pm1.0$ & $324^{+14}_{-17}$      & $ 6.27^{+0.16}_{-0.12}$ \\
WASP-76  & 01:46:31.9 & +02:42:02.0  &  9.42 & $189.0\pm3.0$ & $223.0^{+2.9}_{-2.4}$  & $11.48 \pm 0.07$ \\
WASP-79  & 04:25:29.0 & --30:36:01.6 &  9.98 & $245.7\pm0.9$ & $224.9^{+3.9}_{-3.4}$  & $11.59^{+0.11}_{-0.12}$ \\
WASP-161 & 08:25:21.1 & --11:30:03.6 & 10.90 & $356.2\pm2.0$ & $297.8^{+5.4}_{-12.5}$ & $ 4.80^{+0.12}_{-0.10}$ \\
\hline
\multicolumn{7}{l}{Coordinates were taken from the Gaia Data Release 3 (DR3, Gaia Collaboration \etal 2020).}  \\
\multicolumn{7}{l}{$m_{\rm G}$ is the apparent brightness in the $G$ band from the DR3. Distance is calculated on the}  \\
\multicolumn{7}{l}{DR3 parallaxes with uncertainties from error propagation. $d_{\rm tr}$ and $\delta_{\rm tr}$ are the transit duration}  \\
\multicolumn{7}{l}{and transit depth refined in this study. ppth stands for parts per thousand of the normalised}  \\
\multicolumn{7}{l}{out-of-transit flux.}  \\
}

%%%%% TABLE 2

\MakeTable{lccccc}{12.5cm}{Physical properties of the host stars and their planets of the sample} 
{\hline
Star   & $M_{\star}$ $[M_{\odot}]$   & $R_{\star}$ $[R_{\odot}]$ & $T_{\rm eff}$ [K] & $\log g_{\star}$ [$\rm{cgs}$] & [Fe/H]  \\
\hline 
HAT-P-24  & $1.191 \pm 0.042$      & $1.317 \pm 0.068$         & $6373 \pm 80$  & $4.27 \pm 0.04$ & $-0.16 \pm 0.08$   \\
HAT-P-39  & $1.404 \pm 0.051$      & $1.625^{+0.081}_{-0.062}$ & $6430 \pm 100$ & $4.16 \pm 0.03$ & $+0.19 \pm 0.10$   \\
HAT-P-42  & $1.178 \pm 0.068$      & $1.53 \pm 0.14$           & $5743 \pm 50$  & $4.14 \pm 0.07$ & $+0.27 \pm 0.08$   \\
HAT-P-50  & $1.273^{+0.049}_{-0.115}$ & $1.698 \pm 0.071$      & $6280 \pm 49$  & $4.072 \pm 0.029$ & $-0.18 \pm 0.08$ \\
KELT-2A   & $1.314^{+0.063}_{-0.060}$ & $1.836^{+0.066}_{-0.046}$ & $6148 \pm 48$ & $4.030^{+0.015}_{-0.026}$ & $+0.03\pm0.08$   \\
KELT-15   & $1.181^{+0.050}_{-0.051}$ & $1.481^{+0.091}_{-0.041}$ & $6003^{+56}_{-52}$ & $4.168^{+0.019}_{-0.044}$ & $+0.05 \pm 0.03$   \\
KELT-17   & $1.635^{+0.066}_{-0.061}$ & $1.645^{+0.060}_{-0.055}$ & $7454 \pm 49$ & $4.220^{+0.022}_{-0.024}$ & $-0.018^{+0.074}_{-0.072}$   \\
WASP-23   & $0.78^{+0.13}_{-0.12}$ & $0.765^{+0.033}_{-0.049}$ & $5150 \pm 100$ & $4.4\pm0.2$ & $-0.05\pm 0.13$ \\
WASP-63   & $1.32 \pm 0.05$ & $1.88^{+0.10}_{-0.06}$ & $5550\pm100$ & $3.9\pm0.1$ & $+0.08 \pm 0.07$ \\
WASP-76A  & $1.46 \pm 0.07$ & $1.73 \pm 0.04$ & $6250 \pm 100$ & $4.4 \pm 0.1$ & $+0.23 \pm 0.10$ \\
WASP-79   & $1.56 \pm 0.09$ & $1.64 \pm 0.08$ & $6600 \pm 100$ & $4.20 \pm 0.15$ & $+0.03 \pm 0.10$ \\
WASP-161  & $1.39 \pm 0.14$ & $1.712^{+0.083}_{-0.072}$ & $6400 \pm 100$ & $4.50 \pm 0.15$ & $+0.16 \pm 0.09$ \\
\hline 
Planet     & $M_{\rm b}$ $[M_{\rm Jup}]$ & $R_{\rm b}$ $[R_{\rm Jup}]$ & \multicolumn{3}{l}{Data source} \\
\hline 
HAT-P-24 b & $0.685 \pm 0.033$ & $1.242 \pm 0.067$         & \multicolumn{3}{l}{Kipping \etal (2010)} \\
HAT-P-39 b & $0.599 \pm 0.099$ & $1.571^{+0.108}_{-0.081}$ & \multicolumn{3}{l}{Hartman \etal (2012)} \\
HAT-P-42 b & $1.044 \pm 0.083$ & $1.28 \pm 0.16$           & \multicolumn{3}{l}{Boisse \etal (2013)} \\
HAT-P-50 b & $1.350 \pm 0.073$ & $1.288 \pm 0.064$         & \multicolumn{3}{l}{Hartman \etal (2015)} \\
KELT-2A b  & $1.524 \pm 0.088$ & $1.290^{+0.064}_{-0.050}$ & \multicolumn{3}{l}{Beatty \etal (2012)} \\
KELT-15 b  & $0.91^{+0.21}_{-0.22}$ & $1.443^{+0.110}_{-0.057}$ & \multicolumn{3}{l}{Rodriguez \etal (2016)} \\
KELT-17 b  & $1.31^{+0.29}_{-0.28}$ & $1.525^{+0.065}_{-0.060}$ & \multicolumn{3}{l}{Zhou \etal (2016)} \\
WASP-23 b  & $0.884^{+0.088}_{-0.099}$ & $0.962^{+0.047}_{-0.056}$ & \multicolumn{3}{l}{Triaud \etal (2011)} \\
WASP-63 b  & $0.38 \pm 0.03$ & $1.43^{+0.10}_{-0.06}$ & \multicolumn{3}{l}{Hellier \etal (2012)} \\
WASP-76A b & $0.92 \pm 0.03$ & $1.83^{+0.06}_{-0.04}$ & \multicolumn{3}{l}{West \etal (2016)} \\
WASP-79 b  & $0.90 \pm 0.09$ & $1.70 \pm 0.11$        & \multicolumn{3}{l}{Smalley \etal (2012)} \\
WASP-161 b & $2.49 \pm 0.21$ & $1.143^{+0.065}_{-0.058}$ & \multicolumn{3}{l}{Barkaoui \etal (2019)} \\
\hline
\multicolumn{6}{l}{$M_{\star}$, $R_{\star}$, $T_{\rm eff}$, $g_{\star}$, and [Fe/H] are the mass, radius, effective temperature, gravitational acceleration, }  \\
\multicolumn{6}{l}{and metallicity for the host star, respectively. $M_{\rm b}$ and $R_{\rm b}$ are the mass and radius of the }  \\
\multicolumn{6}{l}{transiting planet in Jupiter units. The data source refers to the parameters of both the star and } \\
\multicolumn{6}{l}{the planet.} \\
}

\textit{HAT-P-24}. This system comprises an inflated hot Jupiter on a 3.4-day orbit and an F8 dwarf star (Kipping \etal 2010). An initially reported non-zero radial velocity (RV) drift attributed to the line-of-sight acceleration of the systemic barycentre was not confirmed (Knutson \etal 2014; Bonomo \etal 2017). The orbit was found to be aligned with the obliquity $\lambda$, \ie a sky-projected angle between the stellar spin and orbital angular momentum, of $20^{\circ} \pm 16^{\circ}$ (Albrecht \etal 2012). The transit parameters were redetermined by Wang \etal (2013), Wang \etal (2021), Hord \etal (2021), and Kokori \etal (2022). Ivshina \& Winn (2022) used TESS observations from sectors 7 and 33 to refine the transit ephemeris, and Hord \etal (2021) searched for transit-like signals in sector 7 data.     
   
\textit{HAT-P-39}. HAT-P-39~b is an inflated hot Jupiter, which completes an orbit around its F-type sun in 3.5 days (Hartman \etal 2012). The systemic parameters were refined by Wang \etal (2021). Ngo \etal (2016) identified a nearby late-type main sequence stellar companion that could be gravitationally bound to the system.

\textit{HAT-P-42}. In this system, an early G star is orbited by a Jupiter-mass planet within 4.6 days (Boisse \etal 2013). Wang \etal (2021) and Patel \& Espinoza (2022) revised the system parameters. Mallonn \etal (2019) and Ivshina \& Winn (2022), based on TESS data from sectors 8 and 34, refined the transit ephemeris. Furthermore, Hord \etal (2021) found no additional transit signatures in sector 8 observations.

\textit{HAT-P-50}. The host star is an F-type dwarf slightly advanced in its evolution stage on the main sequence (Hartman \etal 2015). The planet's transits are observed every 3.1 days. The follow-up studies are mainly based on TESS observations: Ivshina \& Winn (2022) refined the transit ephemeris, Hord \etal (2021) examined out-of-transit photometry for additional transit-like flux drops, and Patel \& Espinoza (2022) and Kokori \etal (2022) redetermined the systemic parameters. 

\textit{KELT-2}. This slightly evolved late F-type dwarf is orbited by a typical hot Jupiter within 4.1 days (Beatty \etal 2012). It is also gravitationally bound with an early K dwarf, constituting a binary stellar system. The host star is the primary and referenced as KELT-2A, and the secondary component is KELT-2B. The presence of water vapour in the planet's atmosphere was detected thanks to high-resolution spectroscopy, which also determined the radial projection of the planetary orbital motion (Piskorz \etal 2018).

\textit{KELT-15}. The host star is a G0 dwarf, close to the end of its evolution at the main-sequence stage. The inflated Jupiter mass planet orbits it every 3.3 days (Rodriguez \etal 2016). The systemic parameters were redetermined by Hord \etal (2021), and the transit ephemeris was refined by Edwards \etal (2021), Kokori \etal (2022), and Ivshina \& Winn (2022).

\textit{KELT-17}. With the obliquity of $-115.9^{\circ} \pm 4.1^{\circ}$, the 3.1-day orbit of KELT-17~b is highly misaligned (Zhou \etal 2016). The host star is a rapidly rotating A-type dwarf. Ivshina \& Winn (2022) refined the transit ephemeris. Garai \etal (2022) used space-borne observations from TESS and the Characterising Exoplanets Satellite (CHEOPS) to refine the systemic parameters. Stangret \etal (2022) acquired high-resolution transmission spectra and did not detect any absorption features that could be attributed to the planet's atmosphere. 

\textit{WASP-23}. Transits of WASP-23~b are observed every 2.9 days (Triaud \etal 2011). Nikolov \etal (2013) acquired the planet's broad-band transmission spectrum, refined its physical parameters, and found the spectral distribution flat. The host star is a K-type dwarf, the only one in our sample. The systemic parameters were refined by Hord \etal (2021). Kokori \etal (2022) and Ivshina \& Winn (2022) refined the transit ephemeris.

\textit{WASP-63}. In this system, the low-density Saturn-mass planet needs 4.4 days to orbit its G8-type star that has likely evolved off the main sequence (Hellier \etal 2012). Kilpatrick \etal (2018) used the Hubble Space Telescope to identify a tentative ${\rm H_{2}O}$ absorption feature in the planetary transmission spectrum. Hord \etal (2021) refined the transit parameters, and Ivshina \& Winn (2022) made the transit ephemeris more precise.

\textit{WASP-76}. The planet can be classified as an extremely hot Jupiter. It moves along a tight orbit with a period of just 1.8 days around an F7-type dwarf (West \etal 2016). The orbit is misaligned with \mbox{$\lambda = 61^{\circ} \pm 7 ^{\circ}$} (Ehrenreich \etal 2020). Southworth \etal (2020) refined the systemic properties and Ivshina \& Winn (2022) refined the transit ephemeris. Bohn \etal (2020) confirmed the host star is accompanied by an $0.8 \, {M_{\odot}}$ common proper motion component, referenced as WASP-76B. Thus, we refer to the host star as WASP-76A. The small orbital distance of the planet results in high irradiance and an equilibrium temperature of about 2200~K, bloating the planetary radius substantially. That gives rise to a broad range of studies on the properties of the planetary atmosphere and detection of atomic and molecular species, such as neutral atomic sodium (Seidel \etal 2019, von Essen \etal 2020), ionised calcium (Tabernero \etal 2021, Deibert \etal 2021) and barium (Azevedo Silva \etal 2022), or titanium oxide and water (Edwards \etal 2020). Furthermore, Edwards \etal (2020) reported on hints for an atmospheric thermal inversion, and Kawauchi \etal (2022) proposed that a thermosphere may also exist. Ehrenreich \etal (2020) observed a blue-shifted spectral feature of neutral iron, which was interpreted as a manifestation of winds blowing from the ultra-hot dayside to relatively cold nightside (May \etal 2021). 

\textit{WASP-79}. The low-density Jupiter-mass planet orbits the F5-type host star within 3.7 days (Smalley \etal 2012). Its orbit was classified as nearly polar (Addison \etal 2013, Brown \etal 2017). The systemic parameters were redetermined by Hord \etal (2021), and then the planetary parameters were verified by Patell and Espinoza (2022). Sotzen \etal (2020) used spectroscopic observations acquired with the Hubble Space Telescope to report on a detection of water, then confirmed by Rathcke \etal (2021). Sotzen \etal (2020) also used white-light data from the Spitzer Space Telescope, the ground-based 6.5 m Magellan II Telescope, and TESS in sectors 4 and 5 to refine the transit ephemeris. Ivshina \& Winn (2022) enhanced the timing data set by incorporating TESS observations in sectors 31 and 32, resulting in even more precise transit ephemeris.

\textit{WASP-161}. The planet is a typical hot Jupiter orbiting its F6-type host star in 5.4 days (Barkaoui \etal 2019). Yang \& Chary (2022) reported on tentatively detecting a departure from a transit linear ephemeris, making the system especially worthy of further observations.

%%%%%%%%%%%%%%%%%%%%%%%%%%%%%%%%%%%%%%%%%%%%%%%%%%%%%%%%%%%%%%%%%%%%%%

\section{TESS observations}

To extract light curves from TESS science frames, we followed the procedure detailed in Maciejewski (2020). The observations were acquired in the short cadence (SC) mode with a 2-minute exposure time. We downloaded target pixel files from the exo.MAST portal\footnote{https://exo.mast.stsci.edu}. For WASP-161 observed in Sector 7, only long cadence (LC, 30 minutes) exposures were available. They were extracted from full-frame images using the TESSCut\footnote{https://mast.stsci.edu/tesscut/} tool (Brasseur \etal 2019). We give a summary of the observations used for each system in Table 3.

Using the aperture photometry method, we obtained the light curves using the Lightkurve v1.9 package (Lightkurve Collaboration \etal 2018). They were preprocessed by removing trends and normalising with the Savitzky-Golay filter. Next, we visually reviewed the light curves to eliminate remnant flux ramps and measurements influenced by scattered light.

%%%%%%%% TABLE 3
 
\MakeTable{cccccccc}{12.5cm}{Details on TESS observations analysed in this study} 
{\hline
Sect./ & from--to & $pnr$ & $N_{\rm tr}$ & Sect./ & from--to & $pnr$ & $N_{\rm tr}$ \\
/Mode  &          & [ppth] &              & /Mode  &          & [ppth] &              \\
\hline 
\multicolumn{4}{c}{HAT-P-24}                 & \multicolumn{4}{c}{KELT-15 (cont.)}          \\
 7/SC & 2019-Jan-07--2019-Feb-02 & 2.98 &  7 & 36/SC & 2021-Mar-07--2021-Apr-02 & 2.29 &  6 \\
33/SC & 2020-Dec-17--2021-Jan-13 & 2.96 &  8 & 61/SC & 2023-Jan-18--2023-Feb-12 & 2.07 &  6 \\
44/SC & 2021-Oct-12--2021-Nov-06 & 3.03 &  6 & 62/SC & 2023-Feb-12--2023-Mar-10 & 1.98 &  8 \\
45/SC & 2021-Nov-06--2021-Dec-02 & 2.86 &  6 & \multicolumn{3}{r}{total:}              & 48 \\
46/SC & 2021-Dec-02--2021-Dec-30 & 2.96 &  6 & \multicolumn{4}{c}{KELT-17}                  \\
\multicolumn{3}{r}{total:}              & 33 & 44/SC & 2021-Oct-12--2021-Nov-06 & 0.98 &  8 \\
\multicolumn{4}{c}{HAT-P-39}                 & 45/SC & 2021-Nov-06--2021-Dec-02 & 0.89 &  7 \\
44/SC & 2021-Oct-12--2021-Nov-06 & 4.43 &  7 & 46/SC & 2021-Dec-02--2021-Dec-30 & 0.92 &  8 \\
45/SC & 2021-Nov-06--2021-Dec-02 & 4.44 &  6 & \multicolumn{3}{r}{total:}              & 23 \\
46/SC & 2021-Dec-02--2021-Dec-30 & 4.43 &  8 & \multicolumn{4}{c}{WASP-23}                  \\
\multicolumn{3}{r}{total:}              & 21 &  6/SC & 2018-Dec-11--2019-Jan-07 & 3.55 &  7 \\
\multicolumn{4}{c}{HAT-P-42}                 & 34/SC & 2021-Jan-13--2021-Feb-09 & 3.63 &  8 \\
 8/SC & 2019-Feb-02--2019-Feb-28 & 3.63 &  4 & 61/SC & 2023-Jan-18--2023-Feb-12 & 3.84 &  7 \\
34/SC & 2021-Jan-13--2021-Feb-09 & 3.46 &  6 & \multicolumn{3}{r}{total:}              & 22 \\
44/SC & 2021-Oct-12--2021-Nov-06 & 3.96 &  4 & \multicolumn{4}{c}{WASP-63}                  \\
45/SC & 2021-Nov-06--2021-Dec-02 & 3.65 &  6 &  6/SC & 2018-Dec-11--2019-Jan-07 & 1.84 &  5 \\
46/SC & 2021-Dec-02--2021-Dec-30 & 3.37 &  6 &  7/SC & 2019-Jan-07--2019-Feb-02 & 1.74 &  4 \\
61/SC & 2023-Jan-18--2023-Feb-12 & 3.71 &  5 & 32/SC & 2020-Nov-19--2020-Dec-17 & 1.77 &  6 \\
\multicolumn{3}{r}{total:}              & 31 & 33/SC & 2020-Dec-17--2021-Jan-13 & 1.75 &  5 \\
\multicolumn{4}{c}{HAT-P-50}                 & \multicolumn{3}{r}{total:}              & 20 \\
 7/SC & 2019-Jan-07--2019-Feb-02 & 2.90 &  7 & \multicolumn{4}{c}{WASP-76A}                 \\
34/SC & 2021-Jan-13--2021-Feb-09 & 3.21 &  7 & 30/SC & 2020-Sep-22--2020-Oct-21 & 2.87 & 11 \\
44/SC & 2021-Oct-12--2021-Nov-06 & 2.90 &  8 & 42/SC & 2021-Aug-20--2021-Sep-16 & 2.91 & 11 \\
45/SC & 2021-Nov-06--2021-Dec-02 & 3.01 &  7 & 43/SC & 2021-Sep-16--2021-Oct-12 & 2.64 & 11 \\
46/SC & 2021-Dec-02--2021-Dec-30 & 3.01 &  8 & \multicolumn{3}{r}{total:}              & 33 \\
\multicolumn{3}{r}{total:}              & 37 & \multicolumn{4}{c}{WASP-79}                  \\
\multicolumn{4}{c}{KELT-2A}                  &  4/SC & 2018-Oct-18--2018-Nov-15 & 1.67 &  4 \\
43/SC & 2021-Sep-16--2021-Oct-12 & 0.65 &  6 &  5/SC & 2018-Nov-15--2018-Dec-11 & 1.20 &  7 \\
44/SC & 2021-Oct-12--2021-Nov-06 & 0.62 &  4 & 31/SC & 2020-Oct-21--2020-Nov-19 & 1.24 &  7 \\
45/SC & 2021-Nov-06--2021-Dec-02 & 0.74 &  6 & 32/SC & 2020-Nov-19--2020-Dec-17 & 1.26 &  7 \\
\multicolumn{3}{r}{total:}              & 16 & \multicolumn{3}{r}{total:}              & 25 \\
\multicolumn{4}{c}{KELT-15}                  & \multicolumn{4}{c}{WASP-161}                 \\
 7/SC & 2019-Jan-07--2019-Feb-02 & 1.99 &  6 &  7/LC & 2019-Jan-07--2019-Feb-02 & 1.58 &  4 \\
 9/SC & 2019-Feb-28--2019-Mar-26 & 2.10 &  8 & 34/SC & 2021-Jan-13--2021-Feb-09 & 2.09 &  4 \\
34/SC & 2021-Jan-13--2021-Feb-09 & 2.06 &  8 & 61/SC & 2023-Jan-18--2023-Feb-12 & 1.93 &  4 \\
35/SC & 2021-Feb-09--2021-Mar-07 & 2.13 &  6 & \multicolumn{3}{r}{total:}              & 12 \\
\hline
\multicolumn{8}{l}{Mode specifies long cadence (LC) or short cadence (SC) photometry.}  \\
\multicolumn{8}{l}{$pnr$ is the photometric noise rate in parts per thousand (ppth) of the normalised flux per minute of}     \\
\multicolumn{8}{l}{exposure, see Fulton \etal (2011). $N_{\rm tr}$ is the number of transits qualified for this study.}     \\
}

%%%%%%%%%%%%%%%%%%%%%%%%%%%%%%%%%%%%%%%%%%%%%%%%%%%%%%%%%%%%%%%%%%%%%%

\section{Data analysis and results}

The data analysis follows the procedure described in detail in Maciejewski \etal (2020). Here, we give a brief outline. The TESS photometric time series were modelled with the Transit Analysis Package (TAP, Gazak \etal 2012) to refine transit models and ephemerides for hot Jupiters in the investigated systems. The original light curves were cut into chunks of 5 times the transit duration around the transit midpoints. Only transits with complete photometric coverage were considered. The transit model was coded with the ratio of the planet to star radii $R_{\rm{p}}/R_{\star}$, semi-major axis scaled in star radii $a/R_{\star}$, orbital inclination $i_{\rm{orb}}$, and linear $u_{\rm 1,TESS}$ and quadratic $u_{\rm 2,TESS}$ limb darkening (LD) coefficients of the quadratic law (Kopal 1950). We also determined the mid-transit time ($T_{\rm{mid}}$) for each transit light curve and allowed for any potential photometric trends in the time domain using a second-order polynomial. Flux contamination from nearby stars was accounted for using the flux contamination parameter $c_{\rm F}$ (as defined in Maciejewski \etal 2023). This quantity was calculated from the averaged crowding parameter (CROWDSAP) for each target in each sector, as provided by the TESS data reduction pipeline. We used the Markov Chain Monte Carlo (MCMC) method to determine the most accurate solutions and uncertainties, running 10 chains with $10^6$ steps each. The refined transit parameters are given in Table~4, and the transit models are presented in Fig.~1.

% FIGURE 
\begin{figure}[thb]
\begin{center}
\includegraphics[width=1.0\textwidth]{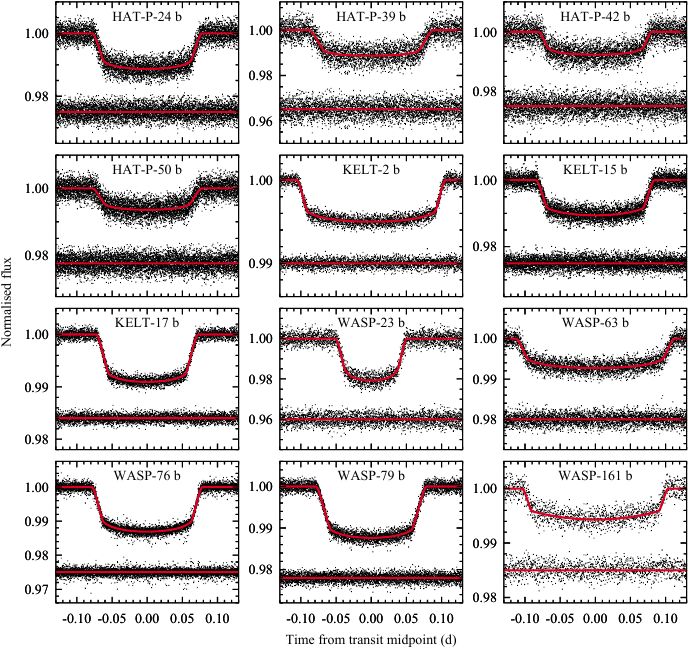}
\end{center}
\FigCap{Phase-folded transit light curves observed with TESS and the best-fitting models for the planets of our sample. The residuals are plotted below each light curve.}
\end{figure}

\MakeTable{lcccccc}{12.5cm}{System parameters from transit light curve modeling}
{\hline
Planet     & $R_{\rm{p}}/R_{\star}$          & $a/R_{\star}$             & $i_{\rm{orb}}$ $[^{\circ}]$ & $u_{\rm 1,TESS}$   & $u_{\rm 2,TESS}$  & $c_{\rm F}$ [\%]\\
\hline
HAT-P-24 b & $0.1007^{+0.0010}_{-0.0011}$    & $7.14^{+0.21}_{-0.18}$    & $87.0^{+0.8}_{-0.6}$    & $0.24^{+0.11}_{-0.10}$ & $0.24^{+0.20}_{-0.21}$ & $2.85$ \\
HAT-P-39 b & $0.1025^{+0.0014}_{-0.0016}$    & $6.14^{+0.29}_{-0.24}$    & $84.84^{+0.95}_{-0.73}$ & $0.20^{+0.23}_{-0.14}$ & $0.19^{+0.27}_{-0.36}$ & $2.65$ \\
HAT-P-42 b & $0.0815^{+0.0011}_{-0.0010}$    & $9.74^{+0.22}_{-0.43}$    & $88.7^{+0.9}_{-1.0}$    & $0.25^{+0.13}_{-0.13}$ & $0.39^{+0.25}_{-0.26}$ & $0.24$ \\
HAT-P-50 b & $0.0788^{+0.0012}_{-0.0013}$    & $5.38^{+0.20}_{-0.17}$    & $82.69^{+0.60}_{-0.54}$ & $0.54^{+0.22}_{-0.30}$ & $-0.07^{+0.40}_{-0.29}$ & $2.37$ \\
KELT-2A b  & $0.0670^{+0.0004}_{-0.0004}$    & $6.30^{+0.15}_{-0.14}$    & $86.84^{+0.72}_{-0.60}$ & $0.33^{+0.07}_{-0.06}$ & $0.08^{+0.11}_{-0.11}$ & $5.50$ \\
KELT-15 b  & $0.0970^{+0.0005}_{-0.0005}$    & $6.74^{+0.11}_{-0.10}$    & $87.64^{+0.64}_{-0.47}$ & $0.37^{+0.06}_{-0.06}$ & $-0.03^{+0.11}_{-0.11}$ & $4.82$ \\
KELT-17 b  & $0.0923^{+0.0005}_{-0.0005}$    & $6.273^{+0.070}_{-0.062}$ & $84.58^{+0.19}_{-0.17}$ & $0.23^{+0.10}_{-0.10}$ & $0.17^{+0.15}_{-0.15}$ & $0.13$ \\
WASP-23 b  & $0.1334^{+0.0019}_{-0.0021}$    & $9.97^{+0.26}_{-0.24}$    & $87.97^{+0.55}_{-0.43}$ & $0.34^{+0.14}_{-0.13}$ & $0.28^{+0.28}_{-0.28}$ & $2.02$ \\
WASP-63 b  & $0.0792^{+0.0010}_{-0.0007}$    & $6.51^{+0.16}_{-0.26}$    & $87.8^{+1.4}_{-1.3}$    & $0.38^{+0.10}_{-0.09}$ & $0.14^{+0.17}_{-0.19}$ & $0.21$ \\
WASP-76A b & $0.10713^{+0.00034}_{-0.00033}$ & $4.064^{+0.033}_{-0.033}$ & $87.5^{+1.0}_{-0.7}$    & $0.306^{+0.026}_{-0.026}$ & $0.157^{+0.054}_{-0.055}$ & $0.03$ \\
WASP-79 b  & $0.10766^{+0.00052}_{-0.00056}$ & $7.148^{+0.074}_{-0.066}$ & $85.53^{+0.18}_{-0.15}$ & $0.32^{+0.09}_{-0.10}$ & $0.05^{+0.15}_{-0.14}$ & $0.77$ \\
WASP-161 b & $0.06931^{+0.00085}_{-0.00074}$ & $8.90^{+0.16}_{-0.37}$    & $90.0^{+0.0}_{-1.8}$    & $0.61^{+0.10}_{-0.13}$ & $-0.28^{+0.22}_{-0.15}$ & $18.2$ \\
\hline
}

We redetermined the mid-transit times for the literature data with TAP to maintain consistency in the transit timing data set. We used only complete light curves publicly available or shared on our request. The transit timing data sets used in the final analysis are presented in Table~5. In the first step, we used those mid-transit times to refine the linear transit ephemerides in the form
\begin{equation}
     T_{\rm mid }(E) = T_0 + P_{\rm orb} \cdot E \, , \;
\end{equation}
where $E$ is the transit number counted from the reference epoch $T_0$, taken from the individual discovery papers. Table~6 presents the results obtained from posterior probability distributions produced by 100 MCMC walkers, each with $10^4$ steps, after discarding the first 1000 steps. These results include the best-fitting parameters and their corresponding $1\sigma$ uncertainties. Figures 2 and 3 illustrate the transit timing residuals plotted against the refined ephemerides.

\MakeTable{ l c c c c l}{12.5cm}{Transit mid-points for hot Jupiters in the studied systems}
{\hline
Planet      & $E$ & $T_{\rm mid}$ [${\rm BJD_{TDB}}$] & $+\sigma$ [d] & $-\sigma$ [d] & Data source\\
\hline
HAT-P-24 b   &   $8$ & $2455243.819418$ & $0.000400$ & $0.000407$ & Kipping \etal (2010)\\
HAT-P-24 b   & $977$ & $2458495.051058$ & $0.000887$ & $0.000902$ & TESS\\
HAT-P-24 b   & $978$ & $2458498.407961$ & $0.000808$ & $0.000822$ & TESS\\
$\cdots$     & $\cdots$ & $\cdots$      & $\cdots$   & $\cdots$   & $\cdots$\\
HAT-P-161 b  & $475$ & $2459984.162116$ & $0.001192$ & $0.001211$ & TESS\\
\hline
\multicolumn{6}{l}{This table is available in its entirety in a machine-readable form at the CDS. }  \\
\multicolumn{6}{l}{A portion is shown here for guidance regarding its form and content.}  \\
}

\MakeTable{ l c c c }{12.5cm}{Transit ephemeris elements for the hot Jupiters in the investigated systems}
{\hline
Planet      & $T_0$ (${\rm BJD_{TDB}}$) & $P_{\rm orb}$ (d) & $\chi^2_{\rm red}$\\
\hline
HAT-P-24 b  & $2455216.97744 \pm 0.00039$ & $3.35524431 \pm 0.00000034$ & 1.15 \\
HAT-P-39 b  & $2455208.7504  \pm 0.0012$  & $3.5438744  \pm 0.0000010$  & 0.81 \\
HAT-P-42 b  & $2455952.5253  \pm 0.0015$  & $4.6418410  \pm 0.0000021$  & 0.95 \\
HAT-P-50 b  & $2456285.91105 \pm 0.00060$ & $3.1220045  \pm 0.0000007$  & 1.13 \\
KELT-2A b   & $2455974.60338 \pm 0.00083$ & $4.1137758  \pm 0.0000010$  & 0.33 \\
KELT-15 b   & $2457029.16979 \pm 0.00037$ & $3.32946620 \pm 0.00000053$ & 1.00 \\
KELT-17 b   & $2457287.74511 \pm 0.00056$ & $3.08018052 \pm 0.00000076$ & 1.11 \\
WASP-23 b   & $2455320.12418 \pm 0.00027$ & $2.94442718 \pm 0.00000022$ & 0.45 \\
WASP-63 b   & $2455921.65314 \pm 0.00063$ & $4.3780821  \pm 0.0000010$  & 0.64 \\
WASP-76A b  & $2456107.85540 \pm 0.00051$ & $1.80988086 \pm 0.00000029$ & 1.62 \\
WASP-79 b   & $2455545.23906 \pm 0.00021$ & $3.6623920  \pm 0.0000003$  & 1.59 \\
WASP-161 b  & $2457416.6210  \pm 0.0010$  & $5.405350   \pm 0.000003$   & 0.28 \\
\hline
\multicolumn{4}{l}{$\chi^2_{\rm red}$ is the reduced chi-square for the refined linear ephemeris}  \\
}

% FIGURE 
\begin{figure}[thb]
\begin{center}
\includegraphics[width=1.0\textwidth]{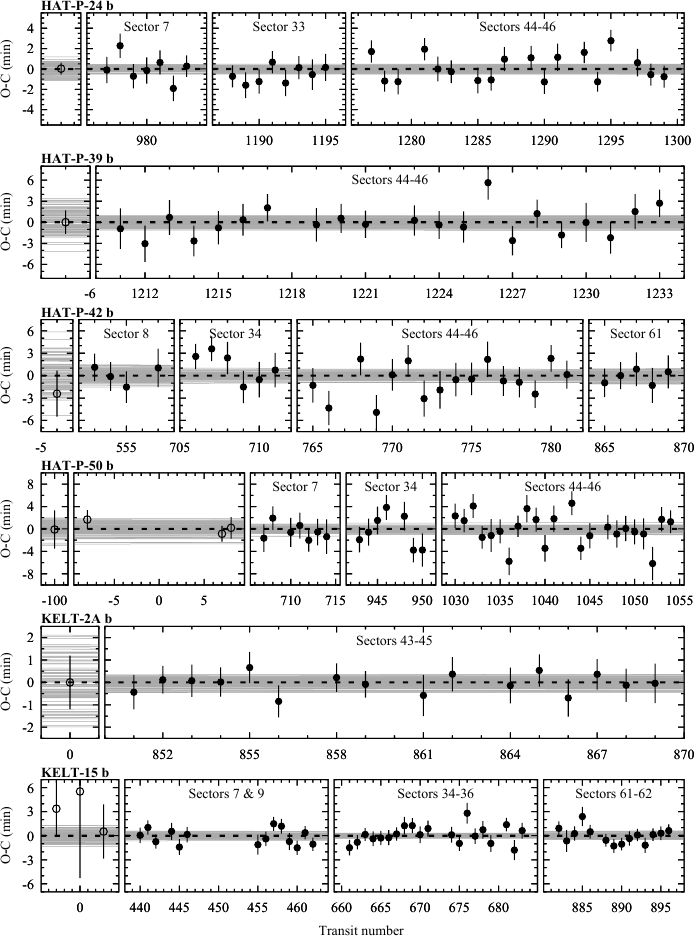}
\end{center}
\FigCap{Transit timing residuals against the refined linear ephemerides for HAT-P-24~b, HAT-P-39~b, HAT-P-42~b, HAT-P-50~b, KELT-2A~b, and KELT-15~b. The mid-transit times from the TESS photometry are marked with filled dots. The redetermined literature values are plotted with open circles. Dashed lines mark the zero value. The ephemeris uncertainties are illustrated by grey lines plotted for 100 parameter sets drawn from the Markov chains.}
\end{figure}

% FIGURE 
\begin{figure}[thb]
\begin{center}
\includegraphics[width=1.0\textwidth]{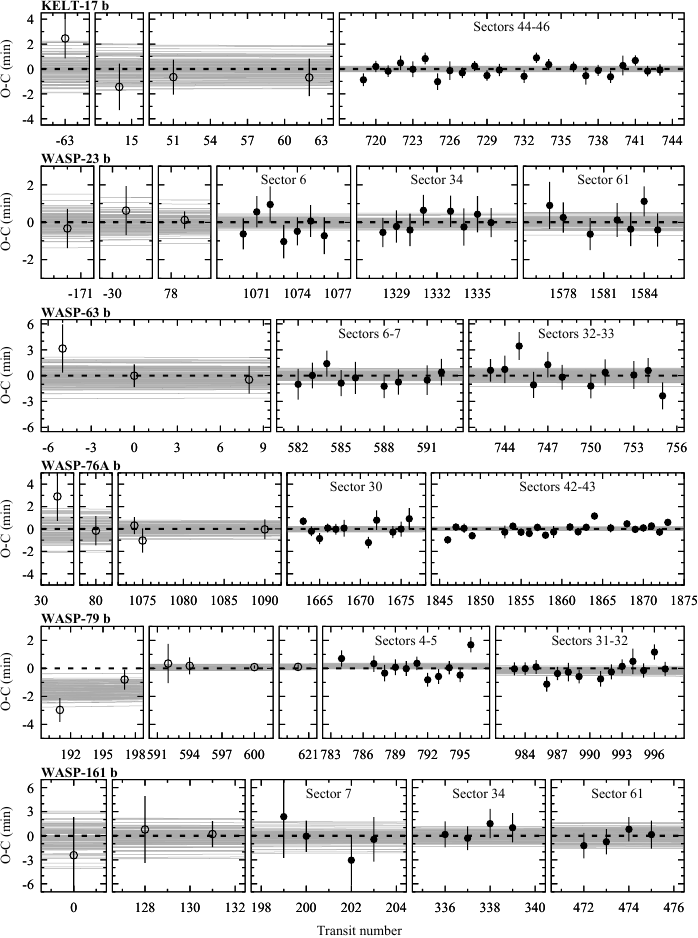}
\end{center}
\FigCap{Same as for Fig.~2, but for KELT-17~b, WASP-23~b, WASP-63~b, WASP-76A~b, and WASP-161~b. For WASP-79~b, the quadratic trend is plotted after subtracting the linear ephemeris.}
\end{figure}

Then, we investigated for any potential long-term trends originating from a monotonic or long-periodic change of $P_{\rm orb}$. We evaluated trail quadratic ephemerides in the form
\begin{equation}
 T_{\rm{mid}}= T_0 + P_{\rm{orb}} \cdot E + \frac{1}{2} \frac{{\rm d} P_{\rm{orb}}}{{\rm d} E} \cdot E^2 \, , \;
\end{equation}
where ${{\rm d} P_{\rm{orb}}}/{{\rm d} E}$ is the change in the orbital period between succeeding transits. Model selection was based on the Bayesian approach, in which the Bayesian information criterion (BIC) values were compared. For 11 planets, constant period scenarios were favoured. For WASP-79~b, the quadratic ephemeris is favoured with $\Delta \rm{BIC} = {\rm{BIC}}_{\rm{linear}} - {\rm{BIC}}_{\rm{quadratic}} \approx 8.3$ and ${{\rm d} P_{\rm{orb}}}/{{\rm d} E} = (-8.0 \pm 2.4) \cdot 10^{-9}$ days per orbital cycle. This solution is plotted in Fig.~3 after subtracting the linear ephemeris, and its significance is discussed in Section 5.2.

We used the Analysis of Variance algorithm (AoV, Schwarzenberg-Czerny 1996) to search for periodic signals in the transit timing residuals against the linear ephemerides. We calculated periodograms for trial periods ranging from 2 to $10^5$ orbital cycles for each planet. We determined the false alarm probability (FAP) levels using the bootstrap method with $10^5$ trials. However, we detected no statistically significant signals for all planets except for KELT-2A~b. We spotted a signal at $2.4753 \pm 0.0004$ orbital cycles with an amplitude of $32.4 \pm 4.3$ s and a FAP of $0.023$\%. The periodogram and the best-fitting periodic model are plotted in Fig.~4. We discuss the potential meaning of this signal in Section 5.1.

% FIGURE 
\begin{figure}[thb]
\begin{center}
\includegraphics[width=1.0\textwidth]{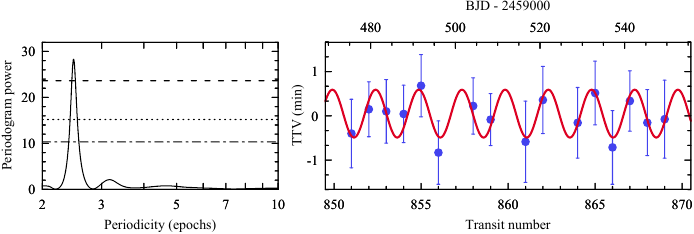}
\end{center}
\FigCap{Left: AoV periodogram produced for transit timing residuals against the linear ephemeris calculated for KELT-2A~b. The Nyquist frequency is 0.5 cycles per epoch (one-half of the sampling rate), corresponding to the period of 2 orbital cycles in the plot. Right: the transit timing residuals for TESS observations (filled dots) with the $\approx 2.5$-cycle periodic model.}
\end{figure}

In the final stage of our analysis, we searched for additional transiting planets with the AoVtr algorithm (Schwarzenberg-Czerny and Beaulieu 2006). We provided it with preprocessed light curves with transit and occultation windows masked out for the hot Jupiters. We tested trial periods ranging from 0.2 to 100 days with a resolution in the frequency domain of $5 \times 10^{-5}$ day$^{-1}$. We iterated over the number of bins from 10 to 100 with a step of 10 to identify the periodogram with the highest peak. We again used the bootstrap method to determine the FAP levels with $10^4$ resampled data sets. As shown in Fig.~5, we did not detect any statistically significant signals (\ie with FAP below 0.1\%) in any of the systems. Next, we conducted injection-recovery tests to establish the transit detection thresholds for the individual systems. We then converted the transit depths into the upper radii of potential planets undetectable in the present photometric time series. The outcomes are presented in Fig.~6.

% FIGURE 
\begin{figure}[thb]
\begin{center}
\includegraphics[width=0.97\textwidth]{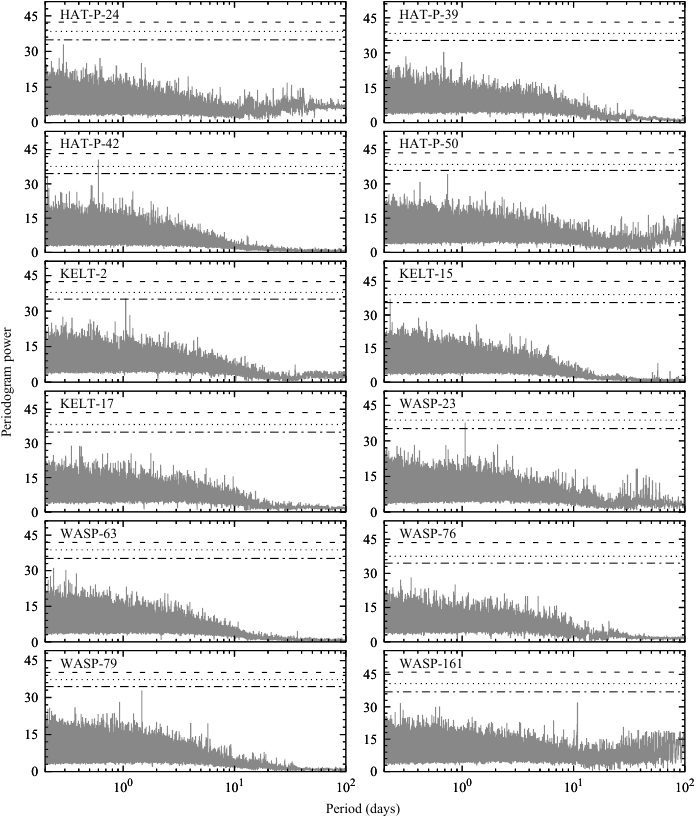}
\end{center}
\FigCap{AoVtr periodograms for out-of-transit observations in the examined systems. The dashed and dotted horizontal lines mark the empirical FAP levels of 5\%, 1\%, and 0.1\% (from the bottom up).}
\end{figure}

% FIGURE 
\begin{figure}[thb]
\begin{center}
\includegraphics[width=0.97\textwidth]{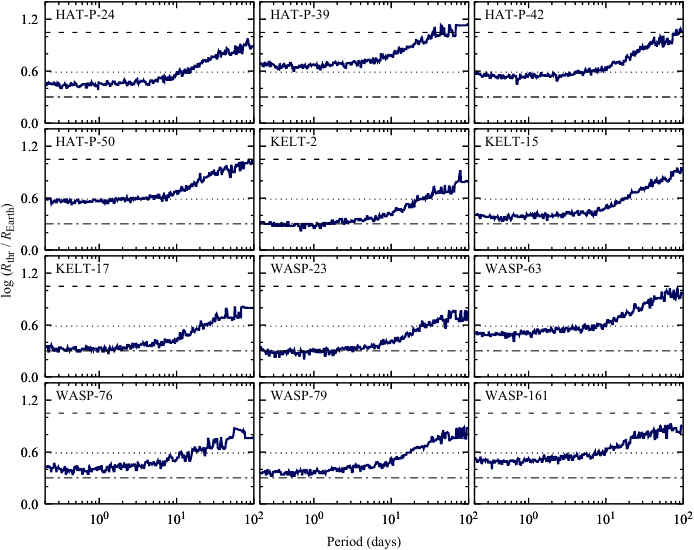}
\end{center}
\FigCap{Empirical upper constraints on radii of hypothetical planets that remain below the transit detection limit in the investigated systems. The dashed and dotted horizontal lines mark the values for 2 $R_{\oplus}$, Neptune, and Jupiter (from the bottom up).}
\end{figure}

%%%%%%%%%%%%%%%%%%%%%%%%%%%%%%%%%%%%%%%%%%%%%%%%%%%%%%%%%%%%%%%%%%%%%%

\section{Discussion}

\subsection{Short-period transit timing variations for KELT-2A b}

The $\approx$2.5-cycle period of the signal detected in transit times for KELT-2A~b might reveal the presence of an exomoon (Kipping 2021). Due to the orbital motion of the parent planet around the planet-moon barycentre, some transits are observed earlier and others later if compared to an unperturbed ephemeris (Sartoretti and Schneider 1999). However, the real frequency of this TTV signal is expected to be greater than the sampling frequency of once every transit (i.e., $1/P_{\rm orb}$). Hence, only harmonic frequencies of the exomoon's orbital period $P_{\rm m}$ can be observed (Kipping 2009). This is the consequence of the 3-body dynamics, for which $P_{\rm m} < P_{\rm orb}$. The TTV signal produced by an exomoon remains undersampled, causing aliasing that prevents the exomoon's orbital period from being determined. Indeed, a periodogram produced for KELT-2A b revealed a set of harmonic frequencies above the Nyquist frequency.

As demonstrated by Kipping (2009), the detection of transit duration variations (TDVs), which are also a consequence of the orbital motion of the planet around the planet-moon barycentre, could help break the degeneracy on $P_{\rm m}$. Compared to the TTV signal, the TDV signal would have the same period and be shifted in phase by $\pi/2$. In a trial iteration, we determined the transit durations for each transit of KELT-2A~b observed by TESS. The periodogram analysis, however, revealed no statistically significant signal. Transit duration uncertainties can explain this negative result, typically an order of magnitude larger than the effect sought.

Assuming the moon's orbit is circular, the dynamically stable configurations reside between the Roche limit and the Hill radius\footnote{In a more conservative approach, the outer limit is given as a fraction of the Hill radius between 1/3 and 1/2.}. For a Moon-like satellite with a density of 3.3 ${\rm g \, cm^{-3}}$, the inner limit falls inside KELT-2A~b due to the planet's low density. The Hill radius is $5.9 \times 10^5 \, {\rm km}$ or $6.6$ planetary radii. For the widest moon's orbit, $P_{\rm m}$ would be 1.7 d, and the mass of the satellite would be 4.7 $M_{\oplus}$ (1\% of the planet's mass). For the tightest orbit, right above the planet, these parameters would be 0.14 d and 24 $M_{\oplus}$ (5\% of the planet's mass), respectively. These calculations show that invoking the exomoon scenario would imply the existence of a super-Earth mass globe orbiting KELT-2A~b. Such massive moons, if they exist, must be extraordinarily rare. Observations show that even less massive Galilean-sized moons are uncommon around planets with semi-major axes between 0.1 and 1 AU (Teachey \etal 2018). On the other hand, the super-Earth satellite around KELT-2A~b would resemble exomoon candidates Kepler-1625~b~I (Teachey \& Kipping 2018), which could be a Neptune-sized globe orbiting a Jupiter-sized planet or Kepler-1708~b~I (Kipping \etal 2022), which could be a 2.6 $R_{\oplus}$ moon around another Jupiter-sized host. We note, however, that those two planets are on orbits much wider than KELT-2A~b (1 and 1.6 AU, respectively).

In the BIC metrics, the sinusoidal signal is not statistically preferred. This finding could come from the overestimated timing uncertainties for individual mid-transit times because the value of the reduced $\chi^2$ for the linear ephemeris was found to be equal to 0.36. After rescaling the uncertainties to satisfy the condition of the reduced $\chi^2$ equal to 1.0, the sinusoidal TTV model was found to be superior to that of the linear ephemeris model with $\Delta {\rm BIC} \approx 5.5$.

\subsection{Transit timing for WASP-79 b}

In the discovery paper, Smalley \etal (2012) used only one incomplete follow-up light curve for WASP-79 b. A closer look revealed a correlated noise at the critical ingress phase, making this photometric time series unsuitable for timing studies. Indeed, our trial analysis resulted in timing uncertainties of 6 minutes. The earliest follow-up light curves were brought by Brown \etal (2017). Then, Sotzen \etal (2020) reported 4 mid-transit times as a by-product of their atmospheric studies. As those photometric time series are publicly unavailable, we used the published determinations with scaled uncertainties. The authors also used timing data from the TESS observations in sectors 4 and 5. Their values agreed with ours with an average difference of 2 s with timing uncertainties lower by a factor of 1.2 on average. We assumed that both methods bring similar results, but for consistency with our timing data set, the timing uncertainties of Sotzen \etal (2020) were inflated accordingly.

The mid-transit times extracted from the light curves of Brown \etal (2017) were observed slightly earlier compared to the updated linear ephemeris. As they are in the early epochs, they favour the quadratic ephemeris. The orbital period derivative was found to be
\begin{equation}
     \dot{P}_{\rm{orb}} = \frac{1}{P_{\rm{orb}}} \frac{{\rm d} P_{\rm{orb}}}{{\rm d} E} = -69 \pm 21 \, {\rm ms \, yr^{-1}}. \;
\end{equation}
This value translates into the barycentre deceleration in line of sight of $\dot{\gamma} = -0.180 \pm 0.054 \, {\rm m \, s^{-1} \, day^{-1}}$. The only useful radial velocity (RV) data set comes from Smalley \etal (2012). It contains 21 measurements that span 422 days. They were acquired with the CORALIE spectrograph on the Swiss 1.2 m telescope between 2010 Dec and 2012 Feb\footnote{We rejected one measurement acquired during a transit affected by the Rossiter-McLaughlin effect.}. We used the Systemic Console Package (Meschiari \etal 2009) to place constraints on $\dot{\gamma}_{\rm RV}$ from the RV data enhanced with the transit timing data set. We found $\dot{\gamma}_{\rm RV} = -0.013 \pm 0.070 \, {\rm m \, s^{-1} \, day^{-1}}$. Both $\dot{\gamma}$ values differ at a $2\sigma$ range, making this result inconclusive. 

It is also worth noting that Sotzen \etal (2020) refined the transit ephemeris and found the reference mid-transit time for epoch 0 from Smalley \etal (2012) to be a $+4.5$ minute outlier. Our test shows that this mid-transit time is de facto a $-4.3$ or $-5.0$ minute outlier, depending on a solution taken: main-sequence or non-main-sequence, respectively. The reason for this discrepancy remains unidentified.

\subsection{Constant orbital period for WASP-161 b}

Yang and Chary (2022) acquired mid-transit times from TESS observations in sectors 7 and 34. They also re-determined a mid-transit time from the only complete ground-based transit light curve acquired with the 1 m robotic SSO-Europa telescope on 2018 Jan 05 and originally reported in the discovery paper by Barkaoui \etal (2019). Other light curves with partial transit coverage were skipped in that study. Interestingly, this SSO-Europa mid-transit time was found to deviate by 7 minutes when compared to predictions of a transit ephemeris based on the TESS data. Furthermore, the reference mid-transit time from Barkaoui \etal (2019) was found to deviate by about 2 hours. A rapid period shortening with $\dot{P}$ of about $-3.7 \; {\rm s \, yr^{-1}}$ was proposed to explain those observations. Similar conclusions were also reported by Shan \etal (2023). 

Our reanalysis showed that the mid-transit times from the Barkaoui \etal (2019) and TESS do follow the linear ephemeris. We notice that a quadratic ephemeris probing for $\dot{P}$ as a free parameter provides the period derivative consistent with zero within a 1$\sigma$ range. To strengthen this finding, we used the partial transit light curves from Barkaoui \etal (2019), observed on 2016 Jan 28 and 2017 Dec 20 with the 60 cm robotic TRAPPIST-South and TRAPPIST-North instruments. The light curve from 2018 Feb 12 was skipped due to its much lower quality. We also incorporated mid-transit times from the LC TESS photometry, the only available data in sector 7. Although those mid-transit times might be affected by lower cadence and exhibit larger uncertainties, they allowed us to analyse the data set similar to that of Yang and Chary (2022).

The origin of the discrepant results of our studies and those of Yang and Chary (2022) remains unclear. The difference in mid-transit time determined for the SSO-Europa light curve by Yang and Chary (2022) and us is about 5 minutes. It can be partially explained by erroneous timestamps conversion from HJD to BJD. Yang and Chary (2022) claim that the difference between both standards reaches $4$ seconds, hence is negligible. We notice, however, that HJD is based on UTC while TESS BJD is based on TDB, producing an offset of about 70 seconds. We compared our BJD(TBD) mid-transit times for TESS data with the values reported by Yang and Chary (2022) as HJD(UTC). They were found to agree with each other well within 1$\sigma$, proving that the literature mid-transit times are, in fact, in BJD(TBD). The more significant contribution could come from incorrectly preprocessed data. The original light curve exhibits high noise with some outlying data points at the ingress phase. Fitting a transit model to the raw data gives a mid-transit time earlier by about 4 minutes. We notice that limiting the light curve to unaffected egress measurements and fitting the template transit model acquired from the TESS data results in a mid-transit time consistent with the linear ephemeris. In our analysis, we removed obvious outliers before transit modelling, which prevented us from receiving the incorrect result.

Finally, we used partial transit light curves from Barkaoui \etal (2019) to support our findings. Since such data might be more susceptible to incorrect removal of instrumental effects, we usually skip them in our studies. For WASP-161~b, we made a derogation from this rule because the magnitude of the investigated effect is much greater than timing uncertainties, even for partial transit data. The light curve from 2016 Jan 28 coincides with the zeroth epoch of the transit ephemeris as given by Barkaoui \etal (2019). The light curve from 2017 Dec 20 is close to the SSO-Europa light curve and could be used to verify our findings. Both photometric time series provided mid-transit times consistent with the constant orbital period model.

Our investigation shows that the transit at epoch 0 was de facto observed about 130 minutes later than reported in Barkaoui \etal (2019) and the original transit ephemeris requires correction (K.~Barkaoui, priv. comm.).

\section{Conclusions}

Our search for the transiting low-mass planetary companions of hot Jupiters in 12 systems revealed no planets larger than 2--4 $R_{\oplus}$. The regime of smaller planets, however, still remains unexplored with the TESS photometry. It could be effectively probed with wider aperture instruments, such as space-borne CHEOPS (Broeg \etal 2013) or PLATO (Rauer \etal 2014).

Our homogenous transit timing analysis showed that the orbital periods of 11 hot Jupiters are stable in long timescales, including WASP-161~b for which the departure from the constant-period model was claimed in the literature. We provide the carefully verified transit ephemeris for that planet. On the other hand, we report a tentative hint of the orbital period change for WASP-79~b. We stress, however, that this finding necessitates further investigation.

For KELT-2A~b, we observe the 2.5-cycle periodic transit timing variation in the TESS data. Since this TTV signal is located in the so-called exomoon corridor (Kipping 2021), it could be attributed to an exomoon orbiting that hot Jupiter. Further high-quality observations are demanded to confirm its existence or even probe for TDVs that could help unravel its nature.

%%%%%%%%%%%%%%%%%%%%%%%%%%%%%%%%%%%%%%%%%%%%%%%%%%%%%%%%%%%%%%%%%%%%%%

\Acknow{We thank Coel Hellier, Julia Victoria Seidel, Monika Lendl, and John Southworth for sharing the follow-up observations used in their studies of the WASP-63 and WASP-76 systems. We also thank David Brown for double-checking the time stamps in the photometric time series published for the WASP-79 system. GM acknowledges the financial support from the National Science Centre, Poland through grant no. 2016/23/B/ST9/00579. This paper includes data collected with the TESS mission, obtained from the MAST data archive at the Space Telescope Science Institute (STScI). Funding for the TESS mission is provided by the NASA Explorer Program. STScI is operated by the Association of Universities for Research in Astronomy, Inc., under NASA contract NAS 5-26555. This research made use of Lightkurve, a Python package for Kepler and TESS data analysis (Lightkurve Collaboration, 2018). This research has made use of the SIMBAD database and the VizieR catalogue access tool, operated at CDS, Strasbourg, France, and NASA's Astrophysics Data System Bibliographic Services.}

%%%%%%%%%%%%%%%%%%%%%%%%%%%%%%%%%%%%%%%%%%%%%%%%%%%%%%%%%%%%%%%%%%%%%%

\end{document}